\documentclass[twocolumn,showpacs,preprintnumbers,amsmath,amssymb,floatfix,prd]{revtex4}
\usepackage{epsfig}
\usepackage{dcolumn}
\begin{document}
\title{Helicity Parton Distributions at a Future Electron-Ion Collider:\\
A Quantitative Appraisal}
\author{Elke C.\ Aschenauer}\email{elke@bnl.gov}
\affiliation{Physics Department, Brookhaven National Laboratory, Upton, NY~11973, USA}
\author{Rodolfo Sassot}\email{sassot@df.uba.ar}
\affiliation{ Departamento de F\'{\i}sica and IFIBA,  Facultad de Ciencias Exactas y Naturales, Universidad de Buenos Aires, Ciudad Universitaria, Pabell\'on\ 1 (1428) Buenos Aires, Argentina}
\author{Marco Stratmann}\email{marco@bnl.gov}
\affiliation{Physics Department, Brookhaven National Laboratory, Upton, NY~11973, USA}

\begin{abstract}
We present a quantitative assessment of the impact a future electron-ion collider will have
on determinations of helicity quark and gluon densities and their contributions to the
proton spin. Our results are obtained by performing a series of global QCD
analyses at next-to-leading order accuracy based on realistic sets of pseudo-data
for the inclusive and semi-inclusive deep-inelastic scattering of longitudinally polarized
electrons and protons at different, conceivable center-of-mass system energies.
\end{abstract}

\pacs{13.88.+e, 13.60.Hb, 13.85.Ni, 12.38.Bx}

\maketitle

\section{Introduction and Motivation}
%
Helicity-dependent parton densities (PDFs) contain the information to what extent
quarks and gluons with a given light-cone momentum fraction $x$ tend to have their spins aligned with
the spin direction of a nucleon in a helicity eigenstate.
The most precise knowledge about these non-perturbative quantities, along with estimates of their
uncertainties, is gathered from comprehensive
global QCD analyses \cite{deFlorian:2008mr,deFlorian:2009vb} to all available 
data taken in spin-dependent deep-inelastic scattering (DIS),
with and without additional identified hadrons in the final-state, and proton-proton collisions. 
Extractions of helicity PDFs are based on the assumption that they factorize from calculable short-distance
partonic scattering cross sections,
which is expected to be a good approximation for processes characterized by a sufficiently large momentum scale
above, say, about $1\div 2\,\mathrm{GeV}$. 
Current analyses \cite{deFlorian:2008mr,deFlorian:2009vb,Blumlein:2010rn,Leader:2010rb}
are performed consistently at next-to-leading order (NLO)
accuracy of QCD for both the scale evolution of helicity PDFs and the relevant hard scattering cross sections.

Apart from being essential for a comprehensive understanding of the partonic structure of 
hadronic matter, helicity PDFs draw much their relevance from their relation to one of the most fundamental
and basic but yet not satisfactorily answered questions in hadronic physics, namely how the spin of a nucleon is 
composed of the spins and orbital angular momenta of quarks and gluons.
The integrals of helicity PDFs over all momentum fractions $x$ (first moments) at a resolution scale $\mu$,
\begin{equation}
\Delta f(\mu) \equiv \int_0^1 \Delta f(x,\mu) \,dx\;\;,
\label{eq:mom}
\end{equation}
provide information about the contribution of a given parton flavor $f$
to the spin of the nucleon.
There are well-known subtleties related to the decomposition of the proton spin in QCD \cite{ref:int-oam},
for instance, $\Delta g(\mu)$ has the interpretation of the gluon spin contribution 
only in light-cone gauge, which is closely tied, however, to the QCD improved parton model,   
but otherwise is a non-local operator. 
The contribution of the quark and antiquark spins to the nucleon spin, 
as summarized by the first moment
of the flavor singlet combination $\Delta\Sigma(\mu)=\sum_{f=q,\bar{q}} \Delta f(\mu)$,
is related to a gauge-invariant operator.
Although orbital angular momenta of quarks and gluons have to be present in the scale evolution
of longitudinally polarized quarks and gluons 
to obey angular momentum conservation in collinear $1\to 2$ parton splittings \cite{Ratcliffe:1987dp}, 
their contributions cannot be quantified from the experimental probes constraining helicity PDFs. 
To this end, information about correlations between the nucleon's spin and transverse degrees of freedom 
of quarks and gluons has to be acquired; see, e.g., Refs.~\cite{ref:int-oam,Boer:2011fh}  
for details.

The relevance of helicity PDFs, and spin physics in general, is also reflected in more than a dozen vigorous
experimental programs in the past twenty-five years, matched by tremendous advancements in
the understanding and development of the underlying theoretical framework.
The most recent global analyses \cite{deFlorian:2008mr,deFlorian:2009vb,Blumlein:2010rn,Leader:2010rb}
confirm early findings that the total quark spin contribution $\Delta \Sigma(\mu)$
is significantly smaller than expectations from naive 
quark models even within still sizable uncertainties from extrapolations
to the unmeasured small $x$ region in Eq.~(\ref{eq:mom}).
A potentially very large gluon spin contribution $\Delta g(\mu) \gtrsim 2$ (in units of $\hbar$), 
initially thought to be a viable way to account for the ``missing'' nucleon spin
\cite{Altarelli:1988nr}, is now strongly disfavored
by measurements of jet and pion yields at BNL-RHIC \cite{ref:rhic}, 
which, on the contrary, prefer a rather small $\Delta g(x,\mu)$
in the range $0.05\lesssim x \lesssim 0.2$ at $\mu\simeq 5\,\mathrm{GeV}$
\cite{deFlorian:2008mr,deFlorian:2009vb,ref:nlo}.
Results for charm and hadron production in polarized lepton-nucleon scattering \cite{ref:fixedtarget} are consistent with
$\Delta g\approx0$ at $x\simeq 0.1$. However, $\Delta g(x,\mu)$ remains to be completely unconstrained at $x\lesssim 0.01$
due to the lack of data and, depending on which functional form one assumes for the extrapolation 
to small values of $x$ in Eq.~(\ref{eq:mom}),
sizable gluon spin contributions of up to $\Delta g(\mu) \simeq 1$ are still conceivable \cite{deFlorian:2008mr,deFlorian:2009vb}.
Theoretical arguments based on the color coherence of gluon couplings suggest
that $\Delta g(x,\mu)\simeq x g(x,\mu)$ in the limit $x\to 0$ at some low but otherwise unspecified bound-state like scale $\mu$ 
\cite{Brodsky:1994kg}. Perturbative evolution to larger scales \cite{ref:qcdevolution}
will change this small $x$ behavior though. While helicity PDFs will not exhibit the strong rise of unpolarized PDFs 
driven by the $1/x$ singularity in the evolution kernel, their actual small $x$ behavior remains unconstrained
by present data. 

The surprisingly small, perhaps even positive strangeness helicity PDF, as 
determined from semi-inclusive DIS (SIDIS) data with identified charged kaons  
in the broad range $0.005\lesssim x \lesssim 0.5$ \cite{deFlorian:2008mr,deFlorian:2009vb,Leader:2010rb},
has triggered quite some discussions recently \cite{Leader:2011tm,ref:dssvplus}. 
If SU(3) flavor symmetry is approximately valid,
one expects a significantly negative first moment of about $\Delta s(\mu)+\Delta\bar{s}(\mu)\simeq -0.1$
by utilizing the experimentally well determined hyperon decay 
constants $F$ and $D$ and the value for $\Delta \Sigma(\mu)$ extracted
from fits to polarized DIS data. 
Recent SIDIS data from COMPASS \cite{ref:compass-sidis} exhibit a weak trend for $\Delta s(x,\mu)$ turning negative somewhere
in the region $0.001\lesssim x\lesssim 0.01$ \cite{ref:dssvplus}, and acquiring a large negative moment $\Delta s(\mu)$
in accordance with SU(3) symmetry is still possible if $\Delta s(x,\mu)$ is large and negative
in the currently unmeasured small $x$ region. However, lattice QCD results for
$\Delta s(\mu)+\Delta\bar{s}(\mu)$ \cite{QCDSF:2011aa} and computations of SU(3) breaking effects in axial current
matrix elements \cite{Savage:1996zd}
point towards sizable violations of SU(3) flavor symmetry, perhaps even consistent
with a vanishing total strangeness polarization $\Delta s(\mu)+\Delta\bar{s}(\mu)\simeq 0$. 
This might be explained by significant chiral corrections as was estimated in \cite{Bass:2009ed}
within the framework of the ``cloudy bag model''.
To complicate things further, all current extractions of strangeness helicity PDFs from SIDIS data exhibit a
significant dependence on the choice of strangeness-to-kaon fragmentation functions (FFs) \cite{ref:compass-sidis}, 
which needs to be scrutinized further. Forthcoming data
from $B$-factories \cite{ref:belleprel},
DIS multiplicities \cite{ref:compassmult}, 
RHIC, and the LHC are likely to considerably improve our knowledge of FFs soon.
Another surprising outcome of analyses of recent COMPASS data \cite{ref:compass-sidis,ref:dssvplus}
was the diminishing evidence for a sizable asymmetry in the light quark
sea, i.e., $\Delta \bar{u}(x,\mu)-\Delta \bar{d}(x,\mu)\neq 0$; 
uncertainties are still large though. Within the large-$N_C$ limit of QCD as incorporated in, e.g.,
the chiral quark soliton model \cite{Diakonov:1996sr}
one expects an SU(2) breaking which is at least as large as what has been
already observed for unpolarized PDFs \cite{Pumplin:2002vw} but with the sign reversed.

Clearly, despite the impressive progress made both experimentally and theoretically 
many fundamental questions related to the proton's helicity structure,
including a quantitative understanding of the decomposition of the proton's spin 
still remain unanswered. The discussions above exemplify the need for measurements
which are sensitive to smaller values of $x$ than accessible 
with past, present, and upcoming polarized DIS fixed-target experiments or with 
high transverse momentum probes at BNL-RHIC. 
An accurate determination of the first moments $\Delta \Sigma(\mu)$ and
$\Delta g(\mu)$ entering the proton's spin sum rule or
elucidating the flavor dependence of helicity PDFs
to quantify, e.g., a potential SU(3) symmetry breaking in the light quark sea
cannot be achieved without considerably enlarging the kinematic coverage 
of spin-dependent data in the future. 
All the required measurements to address and answer these questions 
related to the small $x$ regime are unique to a polarized, high energy lepton-nucleon collider 
such as the proposed electron-ion collider (EIC) project \cite{Boer:2011fh}.

In the remainder of this paper we will demonstrate quantitatively
how polarized DIS and SIDIS measurements at an EIC will improve our
knowledge of helicity quark and gluon densities and their contributions
to the spin of the nucleon.
Our assessment of the impact an EIC is expected to have on the determination
of helicity PDFs is based on a series of
global QCD analyses at NLO accuracy performed with realistic sets of pseudo-data for inclusive
and semi-inclusive DIS measurements at an EIC at various conceivable center-of-mass
system (c.m.s.) energies. Uncertainties of PDFs are estimated with both the robust
Lagrange multiplier method \cite{Stump:2001gu} as well as within the 
Hessian approximation \cite{Pumplin:2001ct}.
We will briefly touch upon related inclusive measurements such as the
Bjorken sum rule, the charm contribution to the polarized structure function
$g_1(x,\mu)$, and novel electroweak probes in spin-dependent DIS.

In the next Section we will describe how the projected data for DIS and SIDIS
are generated, their kinematic coverage, and what kind of cuts 
have been imposed. We also outline our
method to quantify the impact of the EIC data on determinations of helicity PDFs. 
Section III is devoted to detailed discussions of the results of the global
analyses performed with projected EIC data.
Other opportunities related to the nucleon's helicity
structure will be briefly discussed in Sec.~IV.
The main results will be summarized in Section V.

\section{Kinematics, Strategy and Framework}
%
An EIC will most likely be realized in at least two stages with increasing
c.m.s.\ energies \cite{Boer:2011fh}.
To assess the impact of a future EIC in determining helicity PDFs
we will consider two sets of energies conceivable with the first stage of the
eRHIC option of an EIC \cite{ref:erhic} which is based on colliding 
an $E_e=5\,\mathrm{GeV}$ electron
beam with the existing RHIC proton beam of $E_p=100-250\,\mathrm{GeV}$. 
Simulations based on 
pseudo-data generated with an electron energy of $20\,\mathrm{GeV}$ are used
to estimate the impact of a later stage of an EIC.
The resulting c.m.s.\ energies $\sqrt{s}$ 
and corresponding lowest accessible values of $x=Q^2/(s y)$ for two
different values of momentum transfer $Q^2$
are summarized in Tab.~\ref{tab:kine}, assuming
a maximum fractional energy of the virtual photon of $y_{\max}=0.95$.
We only consider c.m.s.\ energies which allow one to access $x$ values
at least down to $10^{-3}$ even for a minimum $Q^2=2.5\,\mathrm{GeV}^2$
to achieve the goal of constraining helicity PDFs in the so far
unexplored small $x$ region.
%
\begin{table}[th!]
\caption{\label{tab:kine} Combinations of electron and proton energies used in our
analyses and the corresponding c.m.s.\ energies and minimum values of $x$
accessible for $Q^2=1$ and $2.5\,\mathrm{GeV}^2$ and $y_{\max}=0.95$.
For each data set a modest integrated luminosity of $10\,\mathrm{fb}^{-1}$ is assumed.}
\begin{ruledtabular}
\begin{tabular}{ccccc}
$E_e\times E_p$   & $\sqrt{s}$       & $x_{\min}$           & $x_{\min}$\\
$[\mathrm{GeV}]$  & $[\mathrm{GeV}]$ & for $Q^2=1\,\mathrm{GeV}^2$ & for $Q^2=2.5\,\mathrm{GeV}^2$ \\ \hline
$5\times 100$     & 44.7             & $5.3\times 10^{-4}$ &  $1.3\times 10^{-3}$ \\
$5\times 250$     & 70.7             & $2.1\times 10^{-4}$ &  $5.3\times 10^{-4}$ \\
$20 \times 250$   & 141.4            & $5.3\times 10^{-5}$ &  $1.3\times 10^{-4}$\\
\end{tabular}
\end{ruledtabular}
\end{table}

%
\begin{figure}[th!]
\begin{center}
\vspace*{-0.6cm}
\epsfig{figure=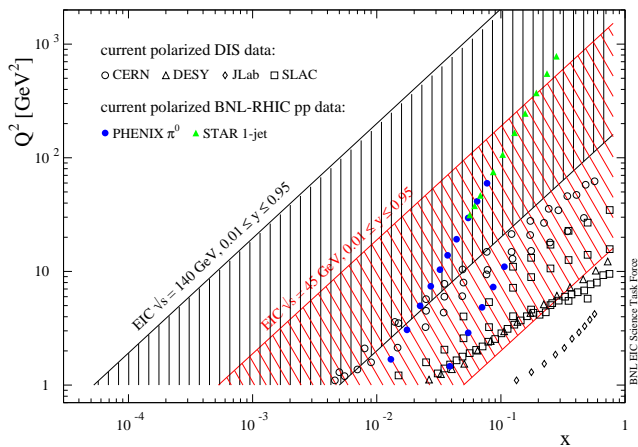,width=0.50\textwidth}
\end{center}
\vspace*{-0.5cm}
\caption{\label{fig:x-q2-plane} [color online]
Kinematic range in $x$ and $Q^2$ accessible with two different c.m.s.\ energies
at an EIC for $0.01\le y \le 0.95$ (hatched areas).
The points illustrate the coverage of currently available data 
from longitudinally polarized DIS and $pp$ experiments (see text).}
\end{figure}
Figure~\ref{fig:x-q2-plane} illustrates the kinematic coverage of an EIC in the $x-Q^2$ plane
for both a first stage with $\sqrt{s}\simeq 45 \,\mathrm{GeV}$ and a full energy eRHIC option
with $\sqrt{s}\simeq 140\,\mathrm{GeV}$.
An EIC will dramatically extend both the $x$ and the $Q^2$ coverage of existing fixed-target DIS experiments,
denoted by the different symbols,
by about two orders of magnitude, opening up unprecedented opportunities for precision studies
of helicity PDFs.
Present QCD analyses \cite{deFlorian:2008mr,deFlorian:2009vb,Blumlein:2010rn,Leader:2010rb}
need to include all DIS data down to $Q^2=1\,\mathrm{GeV}$
in order to reach $x$ values of about $4.6\times10^{-3}$. Allowing for a higher, more
conservative cut in $Q^2$, say $Q^2>2.5\,\mathrm{GeV}^2$, to study the possible relevance of ``higher twist''
corrections, which are suppressed by inverse powers of the hard scale $Q$,
search for possible deviations from standard DGLAP scale evolution \cite{ref:qcdevolution},
or to test the applicability of the assumed factorized pQCD framework, is not possible for the time being
as it would limit the accessible $x$ range too much.
As can be inferred from Tab.~\ref{tab:kine} and Fig.~\ref{fig:x-q2-plane}, at an EIC one can perform all these important studies
and can still reach down to at least $x\simeq 0.001$ even for the lowest c.m.s.\ energy option
we are considering here. 

Also shown in Fig.~\ref{fig:x-q2-plane} are the data from polarized $pp$ collisions at BNL-RHIC \cite{ref:rhic} which
currently provide the best constraint on the polarized gluon distribution. We note that assigning
a transverse momentum dependent observable in $pp$ collisions to a single value of $x$ probed in the gluon
is a gross oversimplification. Choosing $x=2p_T/\sqrt{s}$ nevertheless gives at least a rough idea about the 
lowest $x$ values accessible in $pp$ collisions for a given $p_T$.
Also in DIS at NLO accuracy, information on PDFs is contained within convolutions with hard scattering
cross sections, and $x$ merely reflects the lowest possible momentum fraction accessible in experiment.
The lever-arm in $Q^2$ for any given value of $x$ at an EIC will allow one to extract information on
$\Delta g(x,Q^2)$ from scaling violations, i.e., the rate at which the polarized DIS cross section
changes with scale $Q$ for a given fixed value of $x$. Depending on the lowest value of $Q^2$ used in the analyses,
such studies can be pushed down to about $x\simeq 1\times 10^{-4}$ as will be demonstrated below.

Another unique opportunity at an EIC, although not being pursued further in this study, is to access
novel electroweak structure functions in DIS which probe combinations of quark
helicity PDFs at medium-to-large $x$ and large $Q^2\simeq M_W^2$ different from those constrained
by DIS data at much lower scales and described solely by the exchange of a virtual photon.
Clearly, the kinematic coverage along with the envisioned unprecedented luminosity and the
possibility of having polarized beams are the biggest assets of an EIC which make a suite 
of precision QCD studies possible \cite{Boer:2011fh},
ranging from the proton's helicity structure considered in this paper 
to hadron ``tomography'' through exclusive processes.

To quantify the impact of an EIC on our understanding of helicity PDFs,
we generate sets of pseudo-data for each of the three c.m.s.\ energies listed in Tab.~\ref{tab:kine}.
We use the PEPSI Monte Carlo (MC) generator \cite{Mankiewicz:1991dp} to produce fictitious EIC data for
the inclusive and semi-inclusive DIS of longitudinally polarized electrons and protons
with identified charged pions and kaons in the final-state. We demand a minimum
$Q^2$ of $1\,\mathrm{GeV}^2$, a squared invariant mass of the virtual photon-proton system
larger than $W^2=10\,\mathrm{GeV}^2$, and $0.01\le y\le 0.95$.
The range of $y$ is further restricted from below by constraining the depolarization factor
of the virtual photon 
\begin{equation}
\label{eq:depol}
D(y) = \frac{y(y-2)}{y^2+2(1-y)(1+R)}
\end{equation}
to be larger than $0.1$. $R$ denotes the ratio of the longitudinal to transverse virtual photon
cross sections.
To ensure detection of the scattered lepton we require a minimum momentum of $0.5\,\mathrm{GeV}$,
and, in case of SIDIS, only hadrons with a momentum larger than $1\,\mathrm{GeV}$ 
and a fractional energy in the range $0.2\le z\le 0.9$ are accepted.
All particles detected in the final-state should be at least 1 degree away from the beam directions.
The statistical accuracy of each DIS and SIDIS data set corresponds to a modest 
accumulated integrated luminosity of $10\,\mathrm{fb}^{-1}$, 
equivalent to about one to two months
of operations for the anticipated luminosities for eRHIC \cite{ref:erhic},
except for the $5\times 100\,\mathrm{GeV}$ option which requires about a year of running.

The PEPSI MC \cite{Mankiewicz:1991dp} is currently the only code that allows one
to generate events with definite helicities of
the colliding lepton and proton beams, i.e., to study the longitudinal
double-spin asymmetry 
\begin{eqnarray}
\label{eq:asym}
A_{\parallel} (x,Q^2) &=& \frac{d\sigma^{++}-d\sigma^{+-}}{d\sigma^{++}+d\sigma^{+-}} \\
\label{eq:ratio}
&=& D(y) \frac{g_1(x,Q^2)}{F_1(x,Q^2)}
\end{eqnarray}
which is related to the ratio of virtual photoabsorption cross sections, expressed
by DIS structure functions in (\ref{eq:ratio}), through the depolarization factor $D(y)$.
In (\ref{eq:ratio}), and also in (\ref{eq:depol}), we have neglected kinematic corrections proportional to
$\gamma=\sqrt{4M^2 x^2/Q^2}$, with $M$ the proton mass, which are negligible at
a collider as can be deduced from the $x-Q^2$ coverage depicted Fig.~\ref{fig:x-q2-plane}.
While containing spin-dependent hard scattering matrix elements at LO accuracy, the PEPSI MC
is not capable of simulating parton showers which properly track the polarizations 
of the partons involved, and hence this option has been turned off for generating the EIC data.
QED radiative corrections are known to be sizable and complicate the determination of
the ``true'' values of $x$ and $Q^2$. On the other hand, we have learned a great deal on how
to control and unfold these corrections from years of successful DIS structure function measurements
at DESY-HERA. Undoubtedly, available MC tools \cite{ref:qed} will be further refined in the upcoming years, and
we do not consider QED radiative corrections to be a major limitation on proposed DIS and SIDIS
measurements at an EIC. 
We note that the typical size of the double spin asymmetry (\ref{eq:asym}) at the lowest $x$ values
accessible at an EIC can be as small as a few times $10^{-4}$, depending
on the yet unknown behavior of $\Delta g(x,Q^2)$ in this kinematic regime.
This size sets the scale at which one needs to control systematic uncertainties
due to detector performance or luminosity measurements. Most likely, the dominant source of 
systematic uncertainty will be the determination of the beam polarizations which will lead to a scale uncertainty
in spin asymmetry measurements. 
In any case, DIS measurements,
even with the anticipated high precision, will be far from being the most challenging measurements
to be performed at an EIC \cite{Boer:2011fh}.

Monte Carlo data for the ratio $g_1/F_1$ in DIS and SIDIS 
are generated in 4 [5] bins per decade in $Q^2$ $[x]$ spaced logarithmically.
As the actual pseudo-data used in our global analyses,
we take the ratio $g_1/F_1$ computed at NLO accuracy 
using the DSSV+ \cite{ref:dssvplus} and MRST \cite{Martin:2002aw} polarized and unpolarized PDFs, respectively,
and assign to each $(x,Q^2)$-bin the same relative statistical uncertainties as
obtained with the MC event generator corresponding to
an integrated luminosity of $10\,\mathrm{fb}^{-1}$
and assuming $70\%$ beam polarizations.
In addition, we randomize the pseudo-data in each bin within these one-sigma uncertainties.
In total we add 234 data points for DIS and about 800 points for SIDIS to the 
existing DSSV/DSSV+ global 
analysis framework \cite{deFlorian:2008mr,ref:dssvplus}  based on 570 DIS, SIDIS, and $pp$ data. 
For the SIDIS data with identified charged pions and kaons we assign an additional,
conservative $5$ and $10\%$ relative uncertainty to the EIC pseudo-data to reflect our current
incomplete knowledge of parton-to-pion and parton-to-kaon fragmentation functions, respectively, based
on uncertainty estimates for the DSS sets of FFs 
in Ref.~\cite{deFlorian:2007aj}. 
As already mentioned, various upcoming data sets are expected to greatly advance 
our knowledge of FFs in upcoming years though. If necessary, this can be supplemented
with measurements of unpolarized hadron multiplicities at the EIC. Some expectations
for charged kaon production in unpolarized SIDIS can be found in \cite{Boer:2011fh}.

To quantify the impact of the simulated EIC data on our understanding
of the spin structure of the nucleon, we first need to define some set of reference results
which reflects our current knowledge of helicity PDFs, including a faithful
estimate of their present uncertainties. 
Here, we follow the framework of the original DSSV global analysis \cite{deFlorian:2008mr} and
keep the same functional form, initial scale $Q_0=1\,\mathrm{GeV}$, unpolarized reference PDFs,
standard $\chi^2$ minimization procedure, and number of free fit parameters to facilitate comparisons.
However, as mentioned above, we update the data sets by including
COMPASS DIS \cite{Alekseev:2010hc} and SIDIS \cite{ref:compass-sidis} data which became 
available only after the DSSV analysis was completed.
The resulting new best fit, labeled as DSSV+, which hardly differs from the published DSSV results, is 
used as our baseline fit. 
Despite ongoing experimental efforts, the obtained helicity PDFs will reflect
to a good approximation of what will be known by the time an EIC will start operating.
Some improvements are expected in the meantime at large $x$ and low $Q^2$ from JLab-12 
and towards somewhat smaller values of $x$ than indicated in Fig.~\ref{fig:x-q2-plane} 
from ongoing polarized RHIC $pp$ running at $\sqrt{s}=500\,\mathrm{GeV}$.
Once a sufficient amount of data has been accumulated, measurements of 
single-spin asymmetries in $W$ boson production at RHIC are expected to improve 
$u$ and $d$ quark and antiquark helicity PDFs at $Q\simeq M_W$ and medium-to-large values of $x$ \cite{deFlorian:2010aa} .
As in Ref.~\cite{deFlorian:2008mr}, we will use primarily
the robust Lagrange multiplier method \cite{Stump:2001gu} to quantify uncertainties with and without
including the simulated DIS and SIDIS data. 
In addition, once EIC data are included in the fit, the standard Hessian method \cite{Pumplin:2001ct},
which explores the vicinity of the $\chi^2$ minimum in the quadratic approximation,
also starts to produce reliable results and can be compared to the results obtained with Lagrange
multipliers.

%
%
\begin{figure}[tbh!]
\begin{center}
\vspace*{-0.6cm}
\epsfig{figure=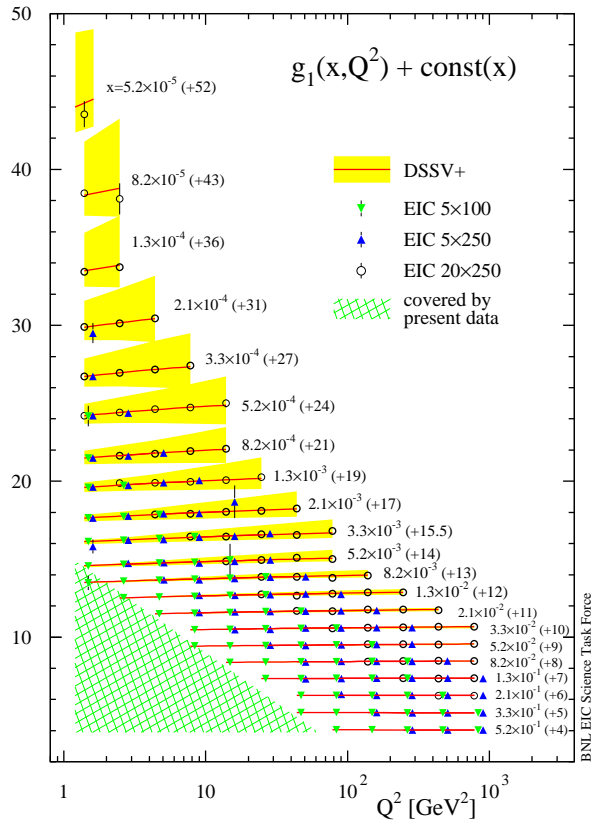,width=0.5\textwidth}
\end{center}
\vspace*{-0.5cm}
\caption{\label{fig:g1} [color online]
Projected EIC data for the structure function $g_1(x,Q^2)$ for three
different combinations of electron and proton energies. Constants are
added to $g_1$ to separate the different $x$ bins. The solid
lines are the result of the DSSV+ best fit, and the shaded bands illustrate
the current uncertainty estimate. Multiple data points at a given
$x, Q^2$ are displaced horizontally to make them more easily visible.
The hatched triangular area indicates the region covered by present data.}
\end{figure}
%
%
\section{Impact of DIS and SIDIS data}
%
Figure~\ref{fig:g1} illustrates our simulated data sets for inclusive polarized DIS at an EIC
for the three different choices of c.m.s.\ energies listed in Tab.~\ref{tab:kine}. 
The error bars were determined as outlined in the previous Section and
reflect the expected statistical accuracy for a modest integrated luminosity of
$10\,\mathrm{fb}^{-1}$.
As indicated by the hatched area, existing fixed target DIS data (see Fig.~\ref{fig:x-q2-plane})
populate only the lower 
left corner of the kinematic plane but connect well or overlap with the lowest $Q^2$ values
accessible with the $5\times 100\,\mathrm{GeV}$ data set.
Relaxing our conservative constraint on the depolarization factor (\ref{eq:depol}),
$D(y)>0.1$, 
would significantly increase the overlap to even lower values of $Q^2$.
We note in passing that if one can control systematic uncertainties very well at an EIC, 
which is definitely the goal, one might try to aim for polarized cross section rather than asymmetry measurements 
in the future.
This would have the added benefit of being independent of the  
ratio $R$ of the longitudinal to transverse virtual photon cross sections.
The shaded bands in Fig.~\ref{fig:g1} correspond to the current uncertainties as 
estimated in the DSSV analysis based on the Lagrange multiplier method.
At low $x$, outside the range constrained by present data, these bands 
essentially reflect the flexibility of the chosen functional form and are
a mere extrapolation.

As is already obvious from Tab.~\ref{tab:kine}, DIS measurements for
$20\times 250\,\mathrm{GeV}$ collisions are crucial to reach $x$ values of
around $10^{-4}$ while still maintaining at least some lever-arm in $Q^2$.
With energies of up to $5\times 250\,\mathrm{GeV}$, envisioned in the first stage
of eRHIC, one can still cover $x$ values down to $5\times 10^{-4}$ for $Q^2\gtrsim 2.5\,\mathrm{GeV}^2$.
Having available an as large as possible range in $Q^2$ for any given fixed value of $x$ 
is of outmost importance for studying scaling violations which are a key prediction of
pQCD. Even though the DIS structure function $g_1$ probes mainly the sum of quark
and antiquark PDFs, its scaling violations at small enough values of $x$ 
are approximately related to the polarized gluon density,
\begin{equation}
\frac{dg_1(x,Q^2)}{d\ln Q^2} \approx - \Delta g(x,Q^2)\,,
\label{eq:scaviol}
\end{equation}
which underlines the importance of precisely measuring them.
In very much the same way, unpolarized DIS data from the DESY-HERA experiments
H1 and ZEUS \cite{Aaron:2009aa} provide the best constraint on
the gluon density at small momentum fractions in all global QCD analyses thanks to
their vast range in $x$ and $Q^2$ only accessible at collider energies.
It is fair to say, that with presently available polarized DIS data one can hardly utilize the relation
(\ref{eq:scaviol}) to determine $\Delta g$ because of the much too limited kinematic coverage.

%
\begin{figure*}[bht!]
\begin{center}
\vspace*{-0.6cm}
\epsfig{figure=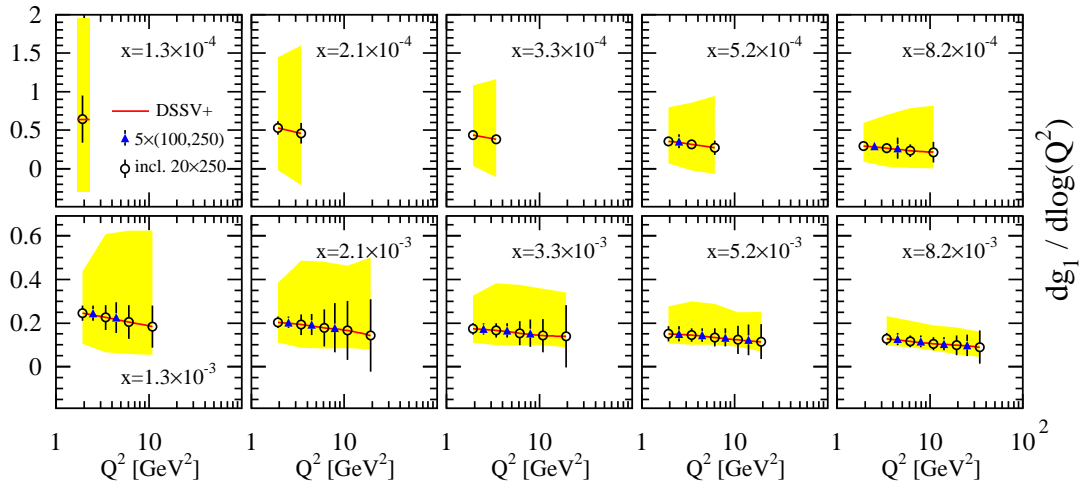,width=0.85\textwidth}
\end{center}
\vspace*{-0.5cm}
\caption{\label{fig:g1scaviol} [color online]
Theoretical expectations for the logarithmic scaling violations
$dg_1(x,Q^2)/d\log Q^2$ based on the DSSV+ best fit (solid lines)
in different bins of $x$. The shaded bands reflect the current
(asymmetric) uncertainties of the DSSV helicity PDFs.
The points illustrate the expected accuracy for measurements at
an EIC with $5\,\mathrm{GeV}$ electrons (triangles) 
based on the projected data shown in Fig.~\ref{fig:g1}.
The open circles include also results obtained with $20\,\mathrm{GeV}$ electrons.
Data points at the same given $(x, Q^2)$ are displaced horizontally to 
distinguish them better.}
\end{figure*}
The prospects for measuring $dg_1(x,Q^2)/d\ln Q^2$ at an EIC are summarized in
Fig.~\ref{fig:g1scaviol}. The projected scaling violations are obtained from the
DIS pseudo-data shown in Fig.~\ref{fig:g1}. For a given bin in $x$, one needs, of course,
at least measurements of $g_1(x,Q^2)$ at two different values of $Q^2$ which are precise enough
to reliably determine the derivative $dg_1(x,Q^2)/d\ln Q^2$ from a difference quotient.
For the binning in $x$ and $Q^2$ adopted in our analysis and the assumed integrated luminosity
of $10\,\mathrm{fb}^{-1}$, a measurement of $dg_1(x,Q^2)/d\ln Q^2$ down to about $x\simeq 1\times 10^{-4}$
appears to be conceivable assuming $20\times 250\,\mathrm{GeV}$ collisions. 
Likewise, a first stage option of an EIC with $5\times 250\,\mathrm{GeV}$
will have sensitivity down to $x\simeq 5\times 10^{-4}$.
This also roughly delineates the range in $x$
where one can expect to put a sensible constraint on $\Delta g(x,Q^2)$ with an inclusive DIS measurement
at an EIC assuming that (\ref{eq:scaviol}) is a good approximation.
The smallness of the projected statistical errors indicates that all inclusive and semi-inclusive
DIS measurements discussed here are systematics limited. Precision measurements will
require a percent-level control of the many different sources of systematic uncertainties such as the
luminosity and polarization measurements but also of the resolution and calibration of the
required detector elements and in the unfolding of QED radiative corrections.

Our projected SIDIS data for identified charged pions and kaons share the same $x$ and $Q^2$ binning
as the DIS data presented in Fig.~\ref{fig:g1} but have slightly larger uncertainties since we assign
an up to $10\%$ additional relative uncertainty due to FFs as explained in Sec.~II.
The SIDIS cross section for a hadron $h$ can be 
expressed by a structure function $g_1^h(x,z,Q^2)$,
$h=\pi^{\pm},K^{\pm}$, which now depends also on the fraction 
$z$ of the momentum of the fragmenting quark or gluon taken by the observed hadron $h$.
$g_1^h(x,z,Q^2)$ exhibits similar scaling violations as the inclusive $g_1$ and will contribute
to constraining $\Delta g$ in a global QCD analysis but SIDIS data draw their relevance from
their sensitivity to different quark and antiquark flavors.
To calibrate measurements of polarized SIDIS at an EIC, one will look at unpolarized
hadron multiplicities first, which on the one hand will help to map out down to which
values of $Q^2$ the leading-twist pQCD framework is a good approximation for SIDIS and
on the other hand will test and improve our knowledge of FFs.

%
%
\begin{figure}[tbh!]
\begin{center}
\vspace*{-0.6cm}
\epsfig{figure=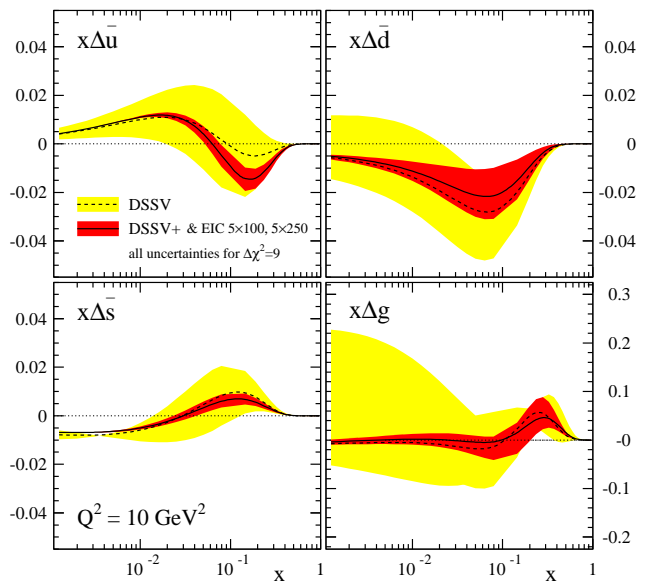,width=0.5\textwidth}
\end{center}
\vspace*{-0.5cm}
\caption{\label{fig:dssz-impact} [color online]
Impact of projected EIC data for the DIS and SIDIS 
of $5\,\mathrm{GeV}$ electrons on $100$ and $250\,\mathrm{GeV}$
protons on the determination of helicity sea quark PDFs and the gluon.
The light shaded bands illustrate present uncertainty estimates 
and the dark shaded, inner bands the improvements expected due to EIC data
(see text).}
\end{figure}
Figure~\ref{fig:dssz-impact} demonstrates the impact of the projected combined 
EIC data for DIS and SIDIS on extractions of the polarized sea quark and gluon densities. 
Here we utilize only data which can be obtained already 
with the initial stage of the eRHIC option, i.e., for collision energies of $5\times 100$ and
$5\times 250\,\mathrm{GeV}$. 
The outer bands in each panel refer to the present ambiguities for helicity PDFs 
as determined in the DSSV analysis \cite{deFlorian:2008mr,deFlorian:2009vb,ref:dssvplus}
and corresponds to a conservative increase in the total $\chi^2$ used to determine the
goodness of the fit by nine units. This value of $\Delta \chi^2$ was regarded to lead to a faithful estimate of 
present uncertainties in \cite{deFlorian:2008mr,deFlorian:2009vb,ref:dssvplus}.
%
%
\begin{figure}[tbh!]
\begin{center}
\vspace*{-0.6cm}
\epsfig{figure=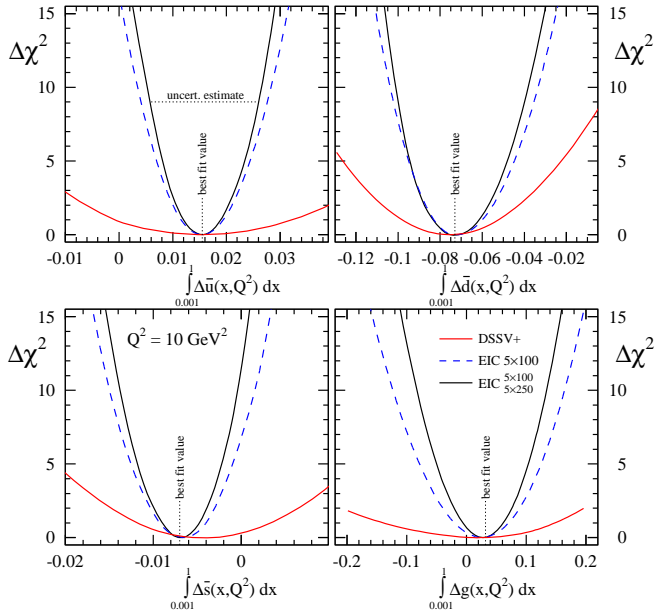,width=0.5\textwidth}
\end{center}
\vspace*{-0.5cm}
\caption{\label{fig:profiles} [color online]
$\chi^2$ profiles for the first moments of helicity sea quark PDFs and the gluon
truncated to the region $0.001\le x \le 1$. The results are based on
using only current data (DSSV+) and sets of projected EIC data with two different
c.m.s.\ energies.}
\end{figure}
The smaller, inner bands are obtained with the same global analysis framework, functional form for the PDFs,
number of free fit parameters, and $\Delta \chi^2$ criterion but now include also the projected EIC data.
As can be seen, the expected improvements are dramatic, in particular, for the 
polarized gluon density below $x\simeq 0.01$ but also for the individual sea quark flavors.
We wish to emphasize that as long as we limit ourselves to the range $x\gtrsim 10^{-3}$, the results shown in
Fig.~\ref{fig:dssz-impact} do not require to analyze data below $Q^2\simeq 2.5\,\mathrm{GeV}$
where the perturbative framework eventually starts to become unreliable and/or where $1/Q$ suppressed power corrections
may become relevant.
At an EIC one can systematically study the validity of the leading twist pQCD framework assumed
in all global QCD analyses so far by varying the lower cut-off scale $Q_{\min}$ above which one 
starts to include data in the fit.
It should also be stressed that only the relative improvement of the uncertainties in Fig.~\ref{fig:dssz-impact},
i.e., the differences between the inner and outer error bands, is of significance here 
for estimating the physics impact of an EIC since the generation of the pseudo-data 
requires to assume a certain set of polarized PDFs.
Of course, only real EIC data will eventually reveal the actual functional form of the helicity PDFs
at small $x$.

%
%
\begin{figure}[tbh!]
\begin{center}
\vspace*{-0.6cm}
\epsfig{figure=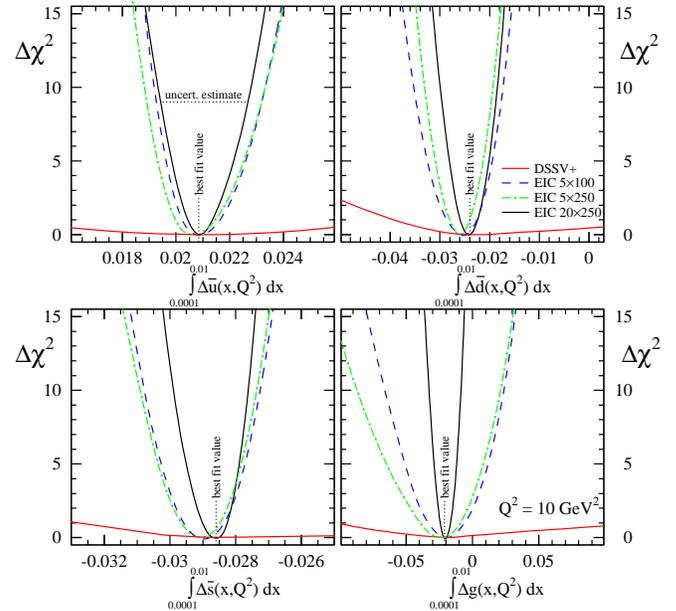,width=0.5\textwidth}
\end{center}
\vspace*{-0.5cm}
\caption{\label{fig:profiles-smallx} [color online]
As in Fig.~\ref{fig:profiles} but now evaluated in the range $0.0001\le x \le 0.01$
and using also a set of projected EIC data for collisions of $20\,\mathrm{GeV}$
electrons on $250\,\mathrm{GeV}$ protons.}
\end{figure}
Figure~\ref{fig:profiles} visualizes the improvements due to EIC data in terms of truncated moments of helicity PDFs, 
which are also used to determine the uncertainty bands in Fig.~\ref{fig:dssz-impact} with the help of the
Lagrange multiplier method \cite{Stump:2001gu,deFlorian:2008mr,deFlorian:2009vb,ref:dssvplus}.
For each parton flavor $f$ we minimize $\chi^2$ with an additional constraint on the value of 
its truncated first moment
\begin{equation}
\Delta f(Q^2,x_{\min},x_{\max})\equiv \int_{x_{\min}}^{x_{max}} \Delta f(x,Q^2) dx
\end{equation}
implemented through a Lagrange multiplier. 
In this way we can map out the $\chi^2$ profile as a function of $\Delta f(Q^2,x_{\min},x_{\max})$ away from its best
fit value without any restrictions on the parameter space. Thereby we generate a large set of alternative PDFs for each point
along the $\chi^2$ contour. Upon choosing a certain maximum increase $\Delta \chi^2$, 
which is still tolerated for a good fit, one
arrives at the uncertainty bands shown in Fig.~\ref{fig:dssz-impact}. 

A more direct way of estimating PDF uncertainties is the standard Hessian method \cite{Pumplin:2001ct}
which is based on the assumption that the $\chi^2$ profiles are quadratic in the vicinity of their minima. 
As can be inferred from the profiles in Fig.~\ref{fig:profiles}, the truncated moments $\Delta f(Q^2,0.001,1)$ are only
very weakly constrained by presently available data, and uncertainties are very large. 
Clearly, the quadratic approximation does only work well for not too large $\Delta \chi^2$ \cite{deFlorian:2009vb}, 
and, hence, reliable Hessian eigenvector PDF sets for $\Delta \chi^2=9$ cannot be constructed to estimate uncertainties.
However, including just one of the projected EIC data sets not only considerably reduces uncertainties for 
$\Delta f(Q^2,0.001,1)$, which can be conveniently read off from the width of the $\chi^2$ profiles
at any desired value of $\Delta \chi^2$, but also leads
to approximately parabolic $\chi^2$ profiles. Hence, to check the consistency of our error estimations, we also
determined the PDF uncertainties with the now applicable Hessian method by constructing appropriate eigenvector PDF sets 
corresponding to $\Delta \chi^2=9$. We find very similar, basically undistinguishable results as for the inner 
uncertainty bands shown in Fig.~\ref{fig:dssz-impact} and obtained with Lagrange multipliers.

It also turns out, see Fig.~\ref{fig:profiles}, that helicity PDFs are already well constrained 
down to $x=1\times 10^{-3}$ by EIC data for $5\times 100\,\mathrm{GeV}$ collision because 
essentially all $x$-bins at that particular c.m.s.\ energy fall into the region $x\gtrsim 10^{-3}$.
Nevertheless, additional data for $5\times 250\,\mathrm{GeV}$ collisions will further 
improve the constraint on $\Delta g$, mainly because of the extended lever-arm in $Q^2$ 
for studies of scaling violations, see Fig.~\ref{fig:g1scaviol}.

To explore the impact of projected EIC data for collisions of $20\,\mathrm{GeV}$
electrons on $250\,\mathrm{GeV}$ protons, envisioned at a full energy eRHIC, 
we perform a similar analysis as in Fig.~\ref{fig:profiles}
but now for the moments truncated in the range from $x=0.0001$ to $0.01$.
As has to be expected, current constraints are even weaker than for the range $x\ge 0.001$
considered in Fig.~\ref{fig:profiles}, resulting
in essentially flat $\chi^2$ profiles. Again, projected EIC data will lead to dramatic improvements even
at an initial stage with only $5\,\mathrm{GeV}$ electrons available. In particular for precision studies of $\Delta g$
at low $x$, the $20\times 250\,\mathrm{GeV}$ data will be extremely crucial as they greatly enlarge the $x$ range where
scaling violations can be studied, see Fig.~\ref{fig:g1scaviol}. This is reflected by the significant
further reduction of the uncertainties in the lower right panel of Fig.~\ref{fig:profiles-smallx}. 
In case that the helicity PDFs
exhibit some sign change at medium-to-large x, as, for instance, $\Delta \bar{u}(x,Q^2)$, $\Delta \bar{s}(x,Q^2)$,
and $\Delta g(x,Q^2)$ in the DSSV analysis \cite{deFlorian:2008mr,deFlorian:2009vb,ref:dssvplus}, 
a numerically significant contribution to their first moments may arise from the small $x$ region, 
i.e., $x\lesssim 0.01$, only accessible at an EIC.   
Having $20\times 250\,\mathrm{GeV}$ data at hand, one can extend the uncertainty bands shown in Fig.~\ref{fig:dssz-impact}
to $x=1\times 10^{-4}$ and perhaps to even lower values of $x$ if data down to $Q^2\simeq 1\div 1.5 \,\mathrm{GeV}^2$ appear to
be amenable to standard leading twist factorization and pQCD methods.

The much reduced uncertainties of helicity PDFs thanks to EIC data will allow one to {\em quantitatively} address most of 
the physics questions concerning the proton's spin structure raised in the Introduction. 
From the $\chi^2$ profiles shown above, one can gather that one can determine the small $x$ 
behavior of $\Delta s(x,Q^2)$ very accurately,
mainly from SIDIS kaon data. In the DSSV analysis \cite{deFlorian:2008mr,deFlorian:2009vb,ref:dssvplus}, 
$\Delta s$ acquires most of its sizable negative $x$-integral 
in the so far unmeasured small $x$ region in order to respect a constraint from SU(3) flavor symmetry. 
The latter is expressed in terms of the two hyperon decay matrix
elements $F$ and $D$ which are experimentally well known, explaining the rather small uncertainty band
for $\Delta s(x,Q^2)$ at small $x$. However, the applicability of this constraint has been questioned,
and large SU(3) breaking effects are certainly not excluded yet \cite{Savage:1996zd,QCDSF:2011aa}.
The ambiguities related to the assumptions about the small $x$ behavior of $\Delta s$, i.e., the
amount of SU(3) breaking, also drives the
current uncertainties of the first moment of the flavor singlet combination 
$\Delta\Sigma$ \cite{deFlorian:2008mr,deFlorian:2009vb,ref:dssvplus} which enters in the proton spin sum rule.

From the $\chi^2$ profile in the lower left panel of Fig.~\ref{fig:profiles-smallx} one can read off that
the truncated moment $\Delta s(Q^2,0.0001,0.01)$ is indeed sizable and negative in the DSSV analysis. 
EIC data are expected to constrain it to within about $5\%$. Clearly, even with modest systematic uncertainties present, 
one can easily quantify the amount of SU(3) flavor breaking effects at an EIC which is a crucial ingredient to 
our understanding of the partonic structure of hadrons and the possible relevance of chiral corrections 
as estimated, for instance, in the cloudy bag model \cite{Bass:2009ed}.
  
Another interesting question related to strangeness is a possible asymmetry $\Delta s(x,Q^2)-\Delta \bar{s}(x,Q^2)\neq 0$
which is also one of the least well determined quantities in case of unpolarized PDFs.
At an EIC, the difference between yields for $K^+$ and $K^-$ will provide sensitivity to such kind of quantities but 
likely requires an improved understanding of the analyzing power given by the ratio of the 
favored to unfavored strangeness fragmentation functions $D_s^{K^{-}}$ and $D_s^{K^+}$, respectively. 
A first feasibility study for unpolarized SIDIS can be found in \cite{Boer:2011fh} but due to the current limitations
for FFs we do not pursue this further here. The LHC is already starting to provide interesting new insights into unpolarized
strangeness distributions from precision measurements of electroweak boson production \cite{ref:atlas}
which can be utilized at the EIC to first improve our knowledge of kaon FFs in unpolarized SIDIS. 
This information should be then sufficient to study $\Delta s(x,Q^2)-\Delta \bar{s}(x,Q^2)$.

%
\begin{figure}[tbh!]
\begin{center}
\vspace*{-0.6cm}
\epsfig{figure=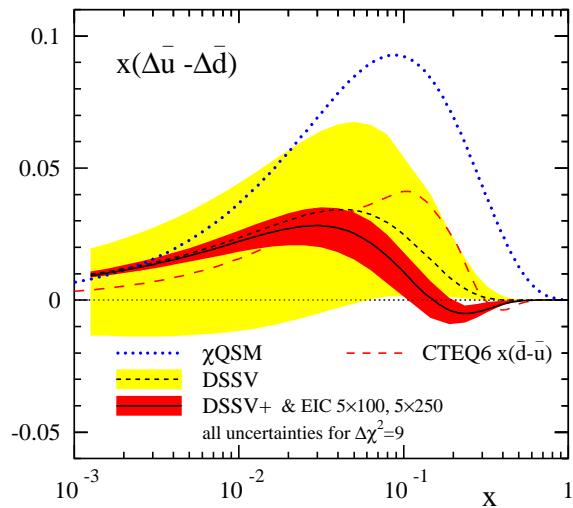,width=0.5\textwidth}
\end{center}
\vspace*{-0.5cm}
\caption{\label{fig:ubar-dbar} [color online] 
As in Fig.~\ref{fig:dssz-impact} but now for the difference of
the light sea quark densities $\Delta \bar{u} - \Delta \bar{d}$.
The dotted line shows an expectation from the chiral quark soliton
model \cite{Diakonov:1996sr} and the dashed line the corresponding asymmetry for
unpolarized PDFs from the CTEQ6 analysis \cite{Pumplin:2002vw} multiplied by $(-1)$.}
\end{figure}
The asymmetry in the light quark sea, $\Delta \bar{u}(x,Q^2)-\Delta \bar{d}(x,Q^2)$, is of particular interest as well.
Firstly, it is know to be sizable in the unpolarized case \cite{Pumplin:2002vw} and secondly it can 
be predicted in various models of
the nucleon structure such as the chiral quark soliton model \cite{Diakonov:1996sr}
where one expects an SU(2) flavor breaking of the sea which is at least as large as what has been already observed 
for unpolarized PDFs but with its sign reversed.
Figure~\ref{fig:ubar-dbar} shows both an expectation from the chiral quark soliton model \cite{Diakonov:1996sr}
and a typical breaking obtained in unpolarized global PDF fits \cite{Pumplin:2002vw}. 
As in Fig.~\ref{fig:dssz-impact}, the larger (outer) error band corresponds to an uncertainty estimate
for $\Delta \bar{u}(x,Q^2)-\Delta \bar{d}(x,Q^2)$ by DSSV based on a Lagrange multiplier analysis of
currently available data and 
a tolerated increase in $\chi^2$ by 9 units. The impact of projected EIC data for $5\times 100$ and
$5\times  250\,\mathrm{GeV}$ collisions is illustrated by the inner error band. This exercise shows that
asymmetries $\Delta \bar{u}(x,Q^2)-\Delta \bar{d}(x,Q^2)$ of about $0.02$ can be resolved, which is 
more than sufficient to test typical model expectations.
We recall that even in the absence of a non-perturbative asymmetry at some low hadronic scale, i.e., 
$\Delta \bar{u}=\Delta \bar{d}$, a non-zero asymmetry will be generated perturbatively through
QCD scale evolution at NLO accuracy \cite{Stratmann:1995fn} and beyond. Likewise, at NNLO an $x$ dependent, local 
$s(x,Q^2)-\bar{s}(x,Q^2)$ asymmetry will develop under QCD scale evolution \cite{Catani:2004nc}.

%
\begin{figure}[tbh!]
\begin{center}
\vspace*{-0.6cm}
\epsfig{figure=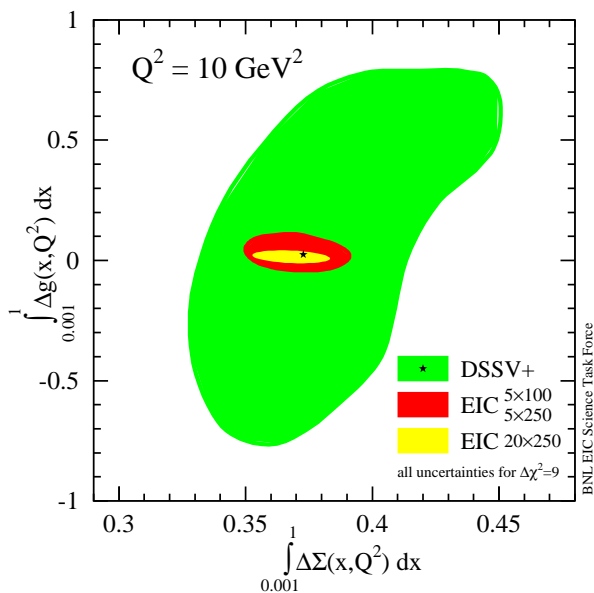,width=0.5\textwidth}
\end{center}
\vspace*{-0.5cm}
\caption{\label{fig:correlation} [color online]
Correlated uncertainties for the first moments of the flavor singlet combination 
$\Delta\Sigma$ and the gluon helicity density $\Delta g$
computed in the region $0.001\le x \le 1$. 
The green, red, and yellow shaded areas are based on fits to current data and to
projected EIC data with $5\,\mathrm{GeV}$ and $20\,\mathrm{GeV}$ electron beams, respectively.
The symbol denotes the DSSV+ best fit.}
\end{figure}
Finally, we look into what can be achieved for the 
first moments of the flavor singlet combination 
$\Delta\Sigma$ and the gluon helicity density $\Delta g$
which both enter the proton spin rule. Figure~\ref{fig:correlation}
shows the correlated uncertainties for the truncated moments
computed in the region $0.001\le x \le 1$ with and without including
projected EIC data sets. As for all our studies presented above, an EIC will greatly
improve the uncertainties, in particular, for $\Delta g$ which is
essentially unconstrained so far. As can be seen,
$\Delta g(Q^2,0.001,1)$ and $\Delta \Sigma(Q^2,0.001,1)$
can be constrained up to about $\pm 0.05$ and $\pm 0.02$, respectively,
if $20\times 250\,\mathrm{GeV}$ data are included in the PDF analyses. However, already at the
initial stage of an EIC a very significant reduction of uncertainties can be achieved.

Again, as a check, very similar uncertainty estimates have been obtained with the Hessian
method once projected EIC data are included in the global analysis framework.
Similar improvements as in Fig.~\ref{fig:correlation} are found for the 
truncated moments for $x>1\times 10^{-4}$ but then $20\times 250\,\mathrm{GeV}$ data
are essential, cf.\ the results for $\Delta g $ shown in Fig.~\ref{fig:profiles-smallx}. 
Although already mentioned in the Introduction, we recall
that the decomposition of the proton spin in Quantum Field Theory is
non-trivial and still under active discussion \cite{ref:int-oam}. While the flavor singlet
contribution $\Delta \Sigma$ appears universally in all proposed spin sum rules, the first moment
of the helicity gluon density acquires the interpretation as the gluon spin
contribution to the proton spin only in light-cone gauge which is the natural
gauge to define parton distributions.
Quark and gluon orbital angular momenta will be part of another
suite of unique measurements at an EIC aiming at the nucleons spatial structure \cite{Boer:2011fh}.

%
\section{Other opportunities in DIS}
%
The presented examples for a physics program with 
longitudinally polarized lepton-nucleon collisions at an EIC are
just the tip of the iceberg and only cover the most important measurements
which can be done in DIS and SIDIS. There are many other avenues at an EIC
which can be pursued to further our understanding of the helicity structure
of the nucleon. For instance, at high $Q^2$ an EIC gives unique access to electroweak effects in
polarized DIS which have not been measured so far. Preliminary studies can be found in \cite{Boer:2011fh}.
Such measurements can be also extended to SIDIS. In general, electroweak DIS results
will constrain helicity PDFs at medium-to-large values of $x$ but at much higher scales
$Q^2\simeq M_W^2$ than currently existing fixed-target DIS data which can be described solely
by one-photon exchange. Although the QCD scale evolution is expected to work well
at large $Q^2$, it has not been tested yet for helicity PDFs. More importantly, 
such measurements should provide a clean way to quantify possible higher twist contributions 
at large momentum fractions $x$ and low $Q^2$ from combined fits.

Another important observable is charm production in polarized DIS, i.e., the charm
contribution $g_1^c(x,Q^2)$ to the inclusive structure function $g_1$. 
So far a proper treatment of heavy flavors in polarized DIS is irrelevant since
in the presently covered $(x,Q^2)$ range its contribution to $g_1$ amounts to less than
$1\%$. At smaller $x$ values, accessible for the first time at an EIC, the size
of $g_1^c(x,Q^2)/g_1(x,Q^2)$ very much depends on what is assumed for the helicity gluon
density as charm is dominantly produced through photon-gluon fusion $\gamma^* g\to c\bar{c}$.
For a small $\Delta g$, as, for instance, in the best fit of DSSV, $g_1^c(x,Q^2)$ remains to be negligible
but can account for as much as $15\%$ of the inclusive $g_1$ at $x=0.001$ for a larger gluon
distribution; see \cite{Boer:2011fh} for some quantitative estimates. 
More theoretical work is clearly needed here, since the relevant cross section
for $\gamma^* g\to c\bar{c}$ with full dependence on the charm quark mass has been
calculated so far only at LO accuracy \cite{Watson:1981ce}. Also, variable flavor number schemes have not been
considered yet for polarized PDF sets and need to be developed in the future.  

It is also conceivable that an EIC can store polarized Helium-3 beams which essentially
act as source of polarized neutrons as $|^3\mathrm{He}\rangle = 0.865 |n\rangle + 2(-0.027) |p\rangle$
and if the spectator protons in an inelastic collision are detected with the help of Roman pots.
The prime physics motivation for studying longitudinally polarized lepton-neutron collisions
is not so much the extra handle on the flavor decomposition but mainly the Bjorken sum
rule \cite{Bjorken:1968dy}
\begin{equation}
\label{eq:bjsum}
\int_0^1 dx\left[ g_1^p(x,Q^2 - g_1^n(x,Q^2)\right] = \frac{1}{6} C_{Bj}[\alpha_s(Q^2)] g_A
\end{equation}
which is currently experimentally verified to about $10\%$.
The Bjorken sum rule is a rare example of a well-understood quantity in pQCD with
${\cal{O}}(\alpha_s^4)$ corrections to $C_{Bj}[\alpha_s(Q^2)]$ being known \cite{Baikov:2010je}
and potentially large $1/Q^2$ higher twist corrections expected to be small in
the perturbative regime \cite{Balitsky:1989jb}.
Of course, significantly improving the current level of experimental accuracy requires, among other things, 
percent level control for $^3\mathrm{He}$ polarimetry which needs some novel technical ideas.
In Ref.~\cite{Boer:2011fh} it has been estimated that one has to access $x$ values down to ${\cal{O}}(10^{-4})$
to limit extrapolation uncertainties in the non-singlet combination 
$\Delta q_3 = \Delta u+\Delta \bar{u} - (\Delta d + \Delta \bar{d})$
effectively probed by the Bjorken sum rule (\ref{eq:bjsum}) to a level of $1-2\%$.
Further theoretical interest in the Bjorken sum is generated from its relation to the Adler
function in $e^+e^-$ annihilation through the Crewther relation \cite{Crewther:1972kn} which has been
worked out up to ${\cal{O}}(\alpha_s^4)$ recently \cite{Baikov:2010je}.
In terms of providing novel information on the flavor separation of helicity PDFs, neutron
data may help to reduce uncertainties for $\Delta d$ and $\Delta \bar{d}$ beyond what
can be achieved with SIDIS data in polarized electron-proton collisions thanks
to the $u \leftrightarrow d$ isospin rotation. 
If technical issues concerning $^3\mathrm{He}$ polarimetry do not prove to be too demanding,
a quantitative estimate of the impact $^3\mathrm{He}$ data on the determination of helicity PDFs
can be made along very similar lines as in our study. Statistical uncertainties will be comparable
to the once obtained for polarized protons in Fig.~\ref{fig:g1}.
 
Finally, precision QCD studies of the helicity structure of nucleons at an EIC may reveal
tensions with DGLAP scale evolution \cite{ref:qcdevolution} which are expected
at sufficiently small but otherwise hard to pinpoint values of $x$ \cite{Bartels:1995iu}.
In contrast to the unpolarized case, the dominant contribution of gluons 
mixes with quarks also at $x\ll 1$. From the standard scale evolution \cite{ref:qcdevolution}
one expects for the small $x$ behavior
\begin{eqnarray}
\nonumber
\Delta q(x,Q^2)\!\!\!&,&\!\!\! \Delta g(x,Q^2) \simeq \\
&&\exp \left[ \mathrm{const}\times
\alpha_s \ln(Q^2/\mu^2) \ln (1/x)\right]^{1/2}
\label{eq:g1smallx}
\end{eqnarray}
assuming for simplicity a fixed coupling $\alpha_s$.
In \cite{Bartels:1995iu} it was demonstrated that this simple behavior
can strongly underestimate the rise at small $x$ due to other 
potentially large double logarithmic contributions
of the type $\alpha_s \ln^2(1/x)^n$ in the $n$-th order of $\alpha_s$ 
which are beyond the standard framework.
This gives rise to a power-like behavior of $g_1$ at small $x$ of the form
$g_1(x,Q^2)\sim(1/x)^{{\cal{O}}(\alpha_s)}$. There are qualitative arguments that
in the polarized case these logarithms in $1/x$ are more relevant than 
in the unpolarized case \cite{Bartels:1995iu}.
Only data can eventually reveal if the kinematic reach of an EIC is sufficient
to actually observe deviations from conventional scale evolution in polarized DIS.

\section{Summary}
%
We have presented a detailed quantitative assessment of the dramatic impact a future EIC will have
on determinations of helicity quark and gluon densities and their contributions to the
proton spin. 
Key asset of a first polarized lepton-nucleon collider will be its 
unprecedented kinematic coverage both down to small momentum fractions $x\simeq 1\times 10^{-4}$ 
and to large scales $Q$, implying a sufficiently large c.m.s.\ energy of the collisions.
This is essential to further our understanding of the nucleon's
helicity structure to level which is sufficient to quantitatively address outstanding
questions about the role of polarized gluons and the flavor structure of sea
quark densities at small momentum fractions $x$.
The necessary precision measurements in polarized inclusive and semi-inclusive deep-inelastic
scattering only require modest integrated luminosities but good control over all sources of systematic uncertainties
ranging from luminosity and polarization measurements, detector acceptance and resolution,
to a proper unfolding of QED radiative corrections.

All presented results were obtained by performing a series of global QCD
analyses at NLO accuracy based on realistic sets of pseudo-data 
for the inclusive and semi-inclusive deep-inelastic scattering of 
longitudinally polarized electrons and protons at different, 
conceivable center-of-mass system energies.
The dramatic physics impact of such data sets has been quantified by
estimating uncertainties for all relevant quantities
with the robust Lagrange multiplier method and by comparing them to
present-day helicity PDF uncertainties.
An EIC will provide precise information on the helicity dependent gluon and flavor separated
quark densities down the momentum fractions of about $10^{-4}$ which in turn will  
accurately determine their contribution to the spin of the proton.  

We have briefly highlighted other interesting opportunities related to helicity PDFs which can
be only pursued at an EIC such as a precision measurement of the Bjorken sum role
which requires, however, to overcome all technical challenges 
related to the need for having an effective polarized neutron beam. 
Charm and electroweak contributions to helicity DIS structure functions
are other prominent examples for measurements uniquely tied to a
high energy polarized lepton-nucleon collider.

\section*{Acknowledgments}
%
E.C.A.\ and M.S.\ are supported by the U.S.\ Department of Energy under contract number DE-AC02-98CH10886.
The work of R.S.\ is supported by CONICET, ANPCyT, and UBACyT.
We acknowledge additional support from a BNL ``Laboratory Directed Research and Development''
grant (LDRD 12-034).


\end{document}